\documentclass{emulateapj}
\usepackage{apjfonts}
\usepackage{graphicx}
\usepackage{amsmath}
\usepackage{natbib}
\usepackage{color}

\lefthead{HENDERSON ET AL.}
\righthead{CANDIDATE GRAVITATIONAL MICROLENSING EVENTS FOR FUTURE DIRECT LENS IMAGING}

\begin{document}

\newcommand{\thetae}{\ensuremath{\theta_{\rm E}}}
\newcommand{\murel}{\ensuremath{\mu_{\rm rel}}}
\newcommand{\te}{\ensuremath{t_{\rm E}}}
\newcommand{\pie}{\ensuremath{\pi_{\rm E}}}
\newcommand{\pirel}{\ensuremath{\pi_{\rm rel}}}
\newcommand{\pivec}{\mbox{\boldmath $\pi$}}
\newcommand{\bdv}[1]{\mbox{\boldmath$#1$}}
\def\bmu{{\bdv{\mu}}}
\def\rel{{\rm rel}}

\title{Candidate Gravitational Microlensing Events for Future Direct Lens Imaging
}

\author{
C. B. Henderson$^{1,43}$,
H. Park$^{2,43}$,
T. Sumi$^{3,44}$,
A. Udalski$^{4,45}$,
A. Gould$^{1,43}$,
Y. Tsapras$^{5,6,46}$,
C. Han$^{2,43,47}$,\\
B. S. Gaudi$^{1,43}$,
V. Bozza$^{7,8}$,\\
and\\
F. Abe$^{9}$,
D. P. Bennett$^{10}$,
I. A. Bond$^{11}$,
C. S. Botzler$^{12}$,
M. Freeman$^{12}$,
A. Fukui$^{13}$,
D. Fukunaga$^{9}$,\\
Y. Itow$^{9}$,
N. Koshimoto$^{3}$,
C. H. Ling$^{11}$,
K. Masuda$^{9}$,
Y. Matsubara$^{9}$,
Y. Muraki$^{9}$,
S. Namba$^{3}$,\\
K. Ohnishi$^{14}$,
N. J. Rattenbury$^{12}$,
To. Saito$^{15}$,
D. J. Sullivan$^{16}$,
D. Suzuki$^{3}$,
W. L. Sweatman$^{11}$,
P. J. Tristram$^{17}$,\\
N. Tsurumi$^{9}$,
K. Wada$^{3}$,
N. Yamai$^{18}$,
P. C. M. Yock$^{12}$
A. Yonehara$^{18}$,\\
(The MOA Collaboration),\\
M. K. Szyma\'nski$^{4}$,
M. Kubiak$^{4}$,
G. Pietrzy\'nski$^{4,19}$,
I. Soszy\'nski$^{4}$,
J. Skowron$^{4}$,
S. Koz{\l}owski$^{4}$,
R. Poleski$^{1,4}$,\\
K. Ulaczyk$^{4}$,
{\L}. Wyrzykowski$^{4,20}$
P. Pietrukowicz$^{4}$,\\
(The OGLE Collaboration),\\
L. A. Almeida$^{21,22}$,
M. Bos$^{23}$,
J.-Y. Choi$^{2}$,
G. W. Christie$^{24}$,
D. L. Depoy$^{25}$,
Subo Dong$^{26}$,
M. Friedmann$^{27}$,
K.-H. Hwang$^{2}$,\\
F. Jablonski$^{28}$,
Y. K. Jung$^{2}$,
S. Kaspi$^{27}$,
C.-U. Lee$^{29}$,
D. Maoz$^{27}$,
J. McCormick$^{30}$,
D. Moorhouse$^{31}$,
T. Natusch$^{24,32}$,\\
H. Ngan$^{24}$,
R. W. Pogge$^{1}$,
I.-G. Shin$^{2}$,
Y. Shvartzvald$^{27}$,
T.-G. Tan$^{33}$,
G. Thornley$^{31}$,
J. C. Yee$^{1,34,35}$\\
(The $\mu$FUN Collaboration),\\
A. Allan$^{36}$,
D. M. Bramich$^{37}$,
P. Browne$^{38}$,
M. Dominik$^{38,39}$,
K. Horne$^{38}$,
M. Hundertmark$^{38}$,
R. Figuera Jaimes$^{38,40}$,\\
N. Kains$^{40}$,
C. Snodgrass$^{41}$,
I. A. Steele$^{42}$,
R. A. Street$^{5}$\\
(The RoboNet Collaboration)
}

\bigskip\bigskip
\affil{$^{1}$Department of Astronomy, The Ohio State University, 140 West 18th Avenue, Columbus, OH 43210, USA}
\affil{$^{2}$Department of Physics, Institute for Astrophysics, Chungbuk National University, Cheongju 371-763, Korea}
\affil{$^{3}$Department of Earth and Space Science, Osaka University, Osaka 560-0043, Japan}
\affil{$^{4}$Warsaw University Observatory, Al. Ujazdowskie 4, 00-478 Warszawa, Poland}
\affil{$^{5}$Las Cumbres Observatory Global Telescope Network, 6740 Cortona Drive, Suite 102, Goleta, CA 93117, USA}
\affil{$^{6}$School of Mathematical Sciences, Queen Mary, University of London, Mile End Road, London E1 4NS, UK}
\affil{$^{7}$Department of Physics, University of Salerno, I-84084-Fisciano (SA), Italy}
\affil{$^{8}$Istituto Nazionale di Fisica Nucleare, Sezione di Napoli, Napoli, Italy}
\affil{$^{9}$Solar-Terrestrial Environment Laboratory, Nagoya University, Nagoya, 464-8601, Japan}
\affil{$^{10}$University of Notre Dame, Department of Physics, 225 Nieuwland Science Hall, Notre Dame, IN 46556-5670, USA}
\affil{$^{11}$Institute of Information and Mathematical Sciences, Massey University, Private Bag 102-904, North Shore Mail Centre, Auckland, New Zealand}
\affil{$^{12}$Department of Physics, University of Auckland, Private Bag 92-019, Auckland 1001, New Zealand}
\affil{$^{13}$School of Chemical and Physical Sciences, Victoria University, Wellington, New Zealand}
\affil{$^{14}$Okayama Astrophysical Observatory, National Astronomical Observatory of Japan, Asakuchi, Okayama 719-0232, Japan}
\affil{$^{15}$Nagano National College of Technology, Nagano 381-8550, Japan}
\affil{$^{16}$Tokyo Metropolitan College of Aeronautics, Tokyo 116-8523, Japan}
\affil{$^{17}$Mt. John University Observatory, P.O. Box 56, Lake Tekapo 8770, New Zealand}
\affil{$^{18}$Department of Physics, Faculty of Science, Kyoto Sangyo University, 603-8555 Kyoto, Japan}
\affil{$^{19}$Universidad de Concepci\'{o}n, Departamento de Astronomia, Casilla 160-C, Concepci\'{o}n, Chile}
\affil{$^{20}$Institute of Astronomy, University of Cambridge, Madingley Road, Cambridge CB3 0HA, UK}
\affil{$^{21}$Instituto de Astronomia, Geof\'{i}sica e Ci\^{e}cias Atmosf\'{e}ricas - IAG/USP Rua do Mat\~{a}o, 1226, Cidade Universit\'{a}ria, S\~{a}o Paulo-SP - Brasil}
\affil{$^{22}$Instituto Nacional de Pesquisas Espaciais, S\~{a}o Jos\'{e} dos Campos, SP, Brazil}
\affil{$^{23}$Molehill Astronomical Observatory, North Shore, New Zealand}
\affil{$^{24}$Auckland Observatory, Auckland, New Zealand}
\affil{$^{25}$Department of Physics and Astronomy, Texas A\&M University, College Station, TX 77843, USA}
\affil{$^{26}$Institute for Advanced Study, Einstein Drive, Princeton, NJ 08540, USA}
\affil{$^{27}$School of Physics and Astronomy, Tel-Aviv University, Tel-Aviv 69978, Israel}
\affil{$^{28}$Observat\'{o}rio do Pico dos Dias, Brasil}
\affil{$^{29}$Korea Astronomy and Space Science Institute, 776 Daedukdae-ro, Yuseong-gu, Daejeon 305-348, Republic of Korea}
\affil{$^{30}$Farm Cove Observatory, Centre for Backyard Astrophysics, Pakuranga, Auckland, New Zealand}
\affil{$^{31}$Kumeu Observatory, Kumeu, New Zealand}
\affil{$^{32}$Institute for Radiophysics and Space Research, AUT University, Auckland, New Zealand}
\affil{$^{33}$Perth Exoplanet Survey Telescope, Perth, Australia}
\affil{$^{34}$Harvard-Smithsonian Center for Astrophysics, 60 Garden St, Cambridge, MA 02138}
\affil{$^{35}$Sagan Fellow}
\affil{$^{36}$School of Physics, University of Exeter Stocker Road, Exeter, Devon, EX4 4QL, UK}
\affil{$^{37}$Qatar Environment and Energy Research Institute, Qatar Foundation, Tornado Tower, Floor 19, P.~O.~Box 5825, Doha, Qatar}
\affil{$^{38}$SUPA, University of St. Andrews, School of Physics and Astronomy, North Haugh, St. Andrews, KY16 9SS, UK}
\affil{$^{39}$Royal Society University Research Fellow}
\affil{$^{40}$European Southern Observatory, Karl-Schwarzschild-Stra{\ss}e 2, 85748 Garching bei M{\"u}nchen, Germany}
\affil{$^{41}$Max Planck Institute for Solar System Research, Max-Planck-Str. 2, 37191 Katlenburg-Lindau, Germany}
\affil{$^{42}$Astrophysics Research Institute, Liverpool John Moores University, Twelve Quays House, Egerton Wharf, Birkenhead, Wirral., CH41 1LD, UK}
\affil{$^{43}$The $\mu$FUN Collaboration}
\affil{$^{44}$The MOA Collaboration}
\affil{$^{45}$The OGLE Collaboration}
\affil{$^{46}$The RoboNet Collaboration}
\affil{$^{47}$Corresponding author}

\begin{abstract}
The mass of the lenses giving rise to Galactic microlensing events can be constrained by measuring the relative lens-source proper motion and lens flux.
The flux of the lens can be separated from that of the source, companions to the source, and unrelated nearby stars with high-resolution images taken when the lens and source are spatially resolved.
For typical ground-based adaptive optics (AO) or space-based observations, this requires either inordinately long time baselines or high relative proper motions.
We provide a list of microlensing events toward the Galactic Bulge with high relative lens-source proper motion that are therefore good candidates for constraining the lens mass with future high-resolution imaging.
We investigate all events from 2004 -- 2013 that display detectable finite-source effects, a feature that allows us to measure the proper motion.
In total, we present 20 events with $\mu \gtrsim$ 8 mas yr$^{-1}$.
Of these, 14 were culled from previous analyses while 6 are new, including OGLE-2004-BLG-368, MOA-2005-BLG-36, OGLE-2012-BLG-0211, OGLE-2012-BLG-0456, MOA-2012-BLG-532, and MOA-2013-BLG-029.
In $\lesssim$12 years from the time of each event the lens and source of each event will be sufficiently separated for ground-based telescopes with AO systems or space telescopes to resolve each component and further characterize the lens system.
Furthermore, for the most recent events, comparison of the lens flux estimates from images taken immediately to those estimated from images taken when the lens and source are resolved can be used to empirically check the robustness of the single-epoch method currently being used to estimate lens masses for many events.
\end{abstract}


\section{Introduction}

Gravitational microlensing provides a useful tool for characterizing Galactic objects in a way that is unbiased by their brightness.
One way to characterize a lens system is to determine its physical parameters by simultaneously measuring the microlens parallax {\pie} and Einstein radius {\thetae}.
The Einstein ring represents the image of the lensed star (source) in the case of exact lens-source alignment, and its radius is commonly used as an angular scale in lensing phenomena.
The Einstein radius is related to the physical parameters of the lens system by ${\thetae} = (\kappa M {\pirel})^{1/2}$, where $M$ is the total mass of the lens system, $\kappa = 4G/(c^{2} {\rm AU})$, ${\pirel} = {\rm AU}(D_{l}^{-1} - D_{s}^{-1})$, and $D_{l}$ and $D_{s}$ are the distances to the lens and source, respectively.
The magnitude of {\pie} corresponds to the relative lens-source parallax, {\pirel}, normalized to {\thetae}.
However, it is generally difficult to measure either {\pie} or {\thetae}, making it all the more unlikely to be able to measure both quantities simultaneously.
As a result, determining the physical quantities of the lens system via this method has been possible only for a small fraction of lensing events.

Another way to characterize a lens is to directly detect the light from the lens system.
In general, direct lens detections are difficult because the typical separations between the lens and source at the time of the event or soon after are on the order of milli-arcseconds (mas), precluding resolution of the lens from the source.
Nevertheless, in principle, measuring the flux of the lens and the source separately can be done without resolving the two systems.
The flux of the source star can be ``de-blended'' from the combined blend flux from all unresolved objects, associated or otherwise, by fitting a microlensing model to the ground-based light curve, since only the source star is magnified during the event.
A high-resolution image of the target then resolves out all unrelated stars with a high probability.
Finally, converting the source flux derived from the ground-based data to the photometric system of the high-resolution data, typically taken in the near-infrared (NIR), and subtracting it from the flux of the target measured in the high-resolution data yields a measurement of excess flux that is not due to the source star itself or to stars with angular separations from the source that are greater than the resolution of the high-resolution data.
Assuming this excess flux is due solely to the lens, the lens mass can then be estimated by combining the resulting mass-distance relation with a mass-luminosity relation and measuring the relative lens-source proper motion (e.g., \citealt{bennett06,bennett07}).

This ``single-epoch'' method has been used to estimate the mass of a number of the hosts of planetary microlensing events (e.g., \citealt{sumi10,batista11}).
However, there exists the possibility that some or all of the excess flux arises from companions to either the lens or the source, or even (with much lower probability) flux from unrelated stars that are blended with the source even in the higher-resolution images.
Furthermore, ground-based $H-$band data exist for only a subset of microlensing events.
For these cases, aligning the ground-based data from which the source flux is derived to the photometric system of the high-resolution data requires a large number of bright and isolated stars, which is difficult to achieve in the ground-based data due to the crowded fields and in the high-resolution image due to the small field of view.
In the case that no ground-based $H-$band data exist, the source flux derived from the ground-based light curve must be transformed from the $I-$band optical flux measurements taken during the event to the NIR flux measurements of the high-resolution data.
Thus, there are significant uncertainties present, whether in the alignment between the two photometric systems or the flux transformation from optical to NIR.

It is possible to circumvent many of the potential difficulties of the single-epoch method by simply waiting until the source and lens have separated sufficiently that they are resolved in a high-resolution image \citep{han03}.
Then, under the assumption that any potential companions to the lens are sufficiently dim, the flux of the lens can be measured directly, thereby eliminating any potential contamination from unrelated stars or companions to the source as well as any need to calibrate the photometry to the ground-based light curve data.
However, most microlensing events toward the Galactic Bulge have sufficiently small relative lens-source proper motions such that resolving the lens and source requires one to wait an inordinately long time.

In this paper, we present a catalog of lensing events toward the Galactic Bulge discovered over the period 2004 -- 2013 with sufficiently high relative proper motions of $\mu \gtrsim 8$ mas yr$^{-1}$ to allow for resolution of the lens and source and thus a direct measurement of the lens flux within $\lesssim$12 years, given the $\sim$0.1$\arcsec$ resolution of ground-based Adaptive Optics (AO) systems and space telescopes.
A subset of these events currently retain modest lens-source separations of $\lesssim$20 mas and thus are unresolvable with present high-resolution systems.
For these events, taking an immediate NIR high-resolution image will allow for a single-epoch estimate of the lens flux.
By comparing this measurement to the direct measurement obtained later with an additional high-resolution image taken when the lens and source have separated enough to be resolved, it will be possible to directly and empirically test for contamination from companions to the source or from unrelated stars and to uncover any systematic errors involved in the flux alignment or transformation.
Our catalog includes 20 total events with relative lens-source proper motions of $\mu \gtrsim$ 8 mas yr$^{-1}$.
Of these, 14 have previously published values of $\mu$ while we present the analysis for 6 new events: OGLE-2004-BLG-368, MOA-2005-BLG-36, OGLE-2012-BLG-0211, OGLE-2012-BLG-0456, MOA-2012-BLG-532, and MOA-2013-BLG-029.

The outline of the paper is as follows.
In Section \ref{sec:events} we discuss the selection criteria for the events and describe the data for each new event.
In Section \ref{sec:analysis} we explain the analysis procedure of the events.
We then combine the results of our analysis with a thorough literature search to provide a catalog of lensing events with high relative proper motion in Section \ref{sec:results}.
We also specifically address the current and future prospects of directly imaging the lenses in our catalog.
In Section \ref{sec:conclusions} we present our conclusions.

\section{Event Selection and Data} \label{sec:events}

The relative lens-source proper motion is determined by $\mu = {\thetae}/{\te}$.
Here {\te} represents the time for the source to cross the Einstein radius.
This Einstein time scale is routinely measured from the analysis of an observed lensing light curve.
The Einstein radius is determined by ${\thetae} = \theta_{*}/\rho$, where $\theta_{*}$ is the angular radius of the source star and $\rho$ is the angular source radius normalized to the Einstein radius.
Both $\theta_{*}$ and $\rho$ are observable in the case of certain events, as discussed below.
Then, the relative proper motion is expressed in terms of these observable quantities by
\begin{equation} \label{eq:ppm}
   \mu = \frac{\theta_{*}}{\rho {\te}}.
\end{equation}

We note that the proper motion in equation (\ref{eq:ppm}) is in the geocentric frame of reference.
However, for our purposes we are interested in the (vector) proper motion in the heliocentric frame $\bmu_{\rm hel}$, which is related to the (vector) geocentric proper motion $\bmu_{\rm geo}$ by \citep{gould04,dong09b},
\begin{equation} 
\bmu_{\rm geo} = \bmu_{\rm hel} - \frac{\pirel}{\rm AU}{\bf v}_{\oplus,\perp},
\end{equation}
where {\pirel} is the relative lens-source parallax and ${\bf v}_{\oplus,\perp}$ is the projected Earth motion at the peak of the event.
This conversion thus requires a measurement of {\pirel}, which we lack for all new events presented here.
Regardless, the difference between the geocentric and heliocentric time scales is typically small and so likely does not significantly
affect our derived results.
Therefore, we will heretofore just refer to the magnitude of the geocentric proper motion as $\mu$.

We establish the following criteria for the selection of our catalog:
\begin{enumerate}
   \item The event must display significant finite-source effects.
   \item There must be multiband photometric data for the event.
   \item The data must be of sufficient quality and coverage to measure {\te} and $\rho$ robustly.
   \item The relative proper motion must be high, specifically the value for $\mu$ must be within one sigma of 8 mas yr$^{-1}$, or higher.
\end{enumerate}

Finite-source effects are necessary to measure $\rho$.
These effects become important when different parts of a source star experience different amounts of magnification, i.e., when there is a significant second derivative of the magnification across the surface of the source star.
In the case of a single lensing mass, this corresponds to an area very near the lens.
Thus, prominent finite-source effects can be observed for high-magnification events in which the source passes very near to or over the lens.
For a lens system composed of two masses, finite-source effects can be observed in the regions around caustics.
Caustics represent the locations in the plane of the source where the magnification of a point-source diverges to infinity.
For a binary lens, there exists a set of 1 -- 3 closed caustic curves, depending on the binary separation.
Then, in the case of a binary lens, finite-source effects can be observed when the source approaches near or crosses a caustic.

We require that there exist multiband data in order to estimate $\theta_{*}$.
For sources on the main sequence it is possible to estimate their color from their magnitude with reasonable precision.
However, for sources that reside on the main sequence turnoff or the subgiant branch, such an estimate is highly uncertain.
In these cases the color must be obtained by direct measurement.
As most microlensing events in current surveys have sources that do not lie on the main sequence, we require multiband photometry in order to measure the color explicitly, which then allows us to estimate the de-reddened color and brightness of the source, from which $\theta_{*}$ can be inferred.

\begin{deluxetable*}{lcccc}
\tablecaption{Coordinates of Events}
\tablewidth{0pt}
\tablehead{
\multicolumn{1}{c}{Event}          &
\multicolumn{1}{c}{R.A. (J2000)}   &
\multicolumn{1}{c}{Dec. (J2000)}   &
\multicolumn{1}{c}{$l$}              &
\multicolumn{1}{c}{$b$}
}
\startdata
OGLE-2004-BLG-368    &   17$^{\rm h}$56$^{\rm m}$14$^{\rm s}$\hskip-2pt.91   &   -29$^{\rm h}$36$^{\rm m}$54$^{\rm s}$\hskip-2pt.2   &    0.60$^{\circ}$    &   -2.34$^{\circ}$   \\
MOA-2005-BLG-36      &   17$^{\rm h}$57$^{\rm m}$23$^{\rm s}$\hskip-2pt.71   &   -27$^{\rm h}$27$^{\rm m}$54$^{\rm s}$\hskip-2pt.1   &    2.59$^{\circ}$    &   -1.48$^{\circ}$   \\  
OGLE-2012-BLG-0211   &   18$^{\rm h}$10$^{\rm m}$10$^{\rm s}$\hskip-2pt.96   &   -25$^{\rm h}$01$^{\rm m}$40$^{\rm s}$\hskip-2pt.2   &    6.12$^{\circ}$    &   -2.78$^{\circ}$   \\
OGLE-2012-BLG-0456   &   17$^{\rm h}$54$^{\rm m}$46$^{\rm s}$\hskip-2pt.69   &   -28$^{\rm h}$59$^{\rm m}$51$^{\rm s}$\hskip-2pt.9   &    0.97$^{\circ}$    &   -1.75$^{\circ}$   \\
MOA-2012-BLG-532     &   17$^{\rm h}$58$^{\rm m}$41$^{\rm s}$\hskip-2pt.13   &   -30$^{\rm h}$02$^{\rm m}$11$^{\rm s}$\hskip-2pt.8   &    0.50$^{\circ}$    &   -3.01$^{\circ}$   \\
MOA-2013-BLG-029     &   17$^{\rm h}$55$^{\rm m}$41$^{\rm s}$\hskip-2pt.06   &   -30$^{\rm h}$40$^{\rm m}$29$^{\rm s}$\hskip-2pt.3   &   -0.38$^{\circ}$    &   -2.77$^{\circ}$
\enddata
\tablecomments{
R.A. and Dec. are taken from the alert page for whichever survey group first alerted the event.
}
\label{tab:coordinates}
\end{deluxetable*}

The requirement of precise and dense coverage of a lensing light curve is important for accurately determining the lensing parameters, particularly $\rho$ and {\te}.

We scour the lists of lensing events discovered during the period 2004 -- 2013 and identify 20 events that satisfy the above criteria.
Among them, 14 events have been previously analyzed, and so we extract their values of {\thetae} and $\mu$ from the literature.
We identify 6 new events that meet the above criteria and do not have extant published analyses: OGLE-2004-BLG-368, MOA-2005-BLG-36, OGLE-2012-BLG-0211, OGLE-2012-BLG-0456, MOA-2012-BLG-532, and MOA-2013-BLG-029.
Table \ref{tab:coordinates} lists the equatorial and Galactic coordinates for each newly analyzed event.

\begin{deluxetable*}{lllll}
\tablecaption{Telescopes}
\tablewidth{0pt}
\tablehead{
\multicolumn{1}{c}{Event}              & 
\multicolumn{1}{c}{MOA}                &
\multicolumn{1}{c}{OGLE}               &
\multicolumn{1}{c}{$\mu$FUN}           &
\multicolumn{1}{c}{RoboNet/MiNDSTEp}
}
\startdata
OGLE-2004-BLG-368    &   MJUO 1.8m   &   LCO 1.3m   &   CTIO 1.3m       &              \\
                     &               &              &   Wise 1.0m       &              \\
                     &               &              &   CTIO Yale 1.0m  &              \\
MOA-2005-BLG-36      &   MJUO 1.8m   &              &   CTIO 1.3m       &              \\
                     &               &              &   Auckland 0.4m   &              \\
OGLE-2012-BLG-0211   &   MJUO 1.8m   &   LCO 1.3m   &   CTIO 1.3m       &              \\
                     &               &              &   FCO 0.36m       &              \\
                     &               &              &   Auckland 0.4m   &              \\
                     &               &              &   PEST 0.3m       &              \\
                     &               &              &   Kumeu 0.36m     &              \\
                     &               &              &   Molehill 0.3m   &              \\
OGLE-2012-BLG-0456   &   MJUO 1.8m   &   LCO 1.3m   &   Wise 1.0m       &              \\
                     &               &              &                   &              \\
MOA-2012-BLG-532     &   MJUO 1.8m   &   LCO 1.3m   &   CTIO 1.3m       &   FTN 2.0m   \\
                     &               &              &   OPD 0.6m        &   LT 2.0m    \\
                     &               &              &   Wise 1.0m       &              \\
MOA-2013-BLG-029     &   MJUO 1.8m   &   LCO 1.3m   &   Auckland 0.4m   &              \\
                     &               &              &   PEST 0.3m       &
\enddata
\tablecomments{
MJUO: Mt.~John University Observatory, New Zealand;
LCO: Las Campanas Observatory, Chile;
CTIO: Cerro Tololo Inter-American Observatory, Chile;
FCO: Farm Cove Observatory, New Zealand;
PEST: Perth Exoplanet Survey Telescope, Australia;
OPD: Observat\'orio do Pico dos Dias, Brazil;
FTN: Faulkes Telescope North, Hawaii;
LT: Liverpool Telescope, La Palma, Spain.
}
\label{tab:telescopes}
\end{deluxetable*}

Each of the events analyzed in this paper is named according to the survey group that alerted (i.e., discovered) the event, followed by the year of discovery, observation field, and a sequence number assigned to that event.
Five of the events presented here were independently discovered by the Optical Gravitational Lensing Experiment (OGLE: \citet{udalski03}) and the Microlensing Observations in Astrophysics collaboration (MOA: \citet{bond01,sumi03}), while one event, MOA-2005-BLG-36, was discovered only by MOA.
We label each event such that the name of the group to first alert the event appears first.
These events were also observed by follow-up groups using networks of telescopes, including the Microlensing Follow-Up Network ($\mu$FUN: \citet{gould06}), and RoboNet \citep{tsapras09}.
All events were observed toward the Galactic Bulge, for which the field designation is BLG.
In Table \ref{tab:telescopes} we list the telescopes used for the observation of each event.

All data were reduced by the respective groups using their own photometric pipelines, most of which employ difference image analysis \citep{alard98,wozniak01,bramich08,albrow09}.
We normalize the flux measurement uncertainties of each data set by first adding a quadratic term to make the cumulative $\chi^{2}$ distribution approximately linear as a function of magnification and subsequently rescaling the uncertainties so that $\chi^{2}$/dof for each data set becomes unity for the best-fit model.

\section{Analysis} \label{sec:analysis}

The analysis of each of the 6 events follows the same procedure.
In the case of a single lens, we search for a best-fit set of lensing parameters starting from a preliminary point-source model.
The lensing parameters include the time of the source's closest approach to the lens, $t_{\rm 0}$, the lens-source separation at that moment normalized to the Einstein radius, $u_{\rm 0}$ (impact parameter), and {\te}.
In order to refine the solution, we then run a Markov Chain Monte Carlo (MCMC) search of parameter space.
In this process, we include $\rho$ as an additional parameter to account for finite-source effects.

In the case of a binary lens, we begin with a grid search across the projected separation of the lens components, $s$, normalized by {\thetae}, the mass ratio of the lens components, $q$, and the source trajectory angle with respect to the binary lens axis, $\alpha$.
At each ($s$, $q$, $\alpha$) grid point we run an MCMC search to find the best-fit combination of the remaining lensing parameters $t_{\rm 0}$, $u_{\rm 0}$, {\te}, and $\rho$.
We iteratively run the grid search to narrow down the parameter space around the global minimum.
Starting near the global minimum identified by the grid search we then run a full MCMC search of parameter space, also allowing the grid parameters to vary, to determine the parameters and uncertainties of the best-fit model that minimizes $\chi^{2}$.

\begin{figure}[h]
\centerline{
\includegraphics[width=9cm]{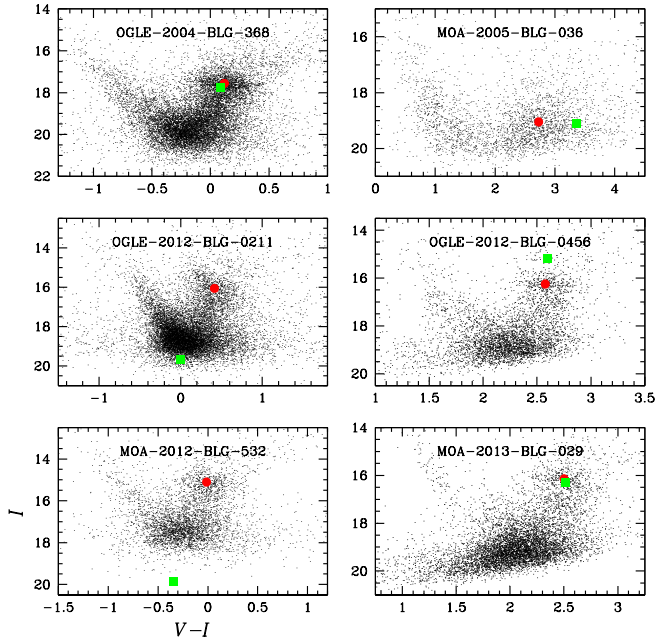}
}
\caption{
The color-magnitude diagrams for all 6 lensing events.
The red filled circle marks the centroid of the giant clump, while the filled green square denotes the source.
The magnitude and color plotted here are instrumental, but we are able to obtain the de-reddened source magnitude and color by using the centroid of the giant clump as a standard candle.
}
\label{fig:cmd}
\end{figure}

For each event, we estimate $\theta_{*}$ based on the multiband ($V$ and $I$) source brightness measured from the analysis of the lensing light curve.
For this, we locate the source star in a color-magnitude diagram of neighboring stars in the same field (see Figure \ref{fig:cmd}).
Using the centroid of the giant clump as a standard candle with $(V-I)_{\rm RC,0} = 1.06$ \citep{bensby11} and accounting for how $I_{\rm RC,0}$ varies with Galactic longitude \citep{nataf13}, we compute the de-reddened $V$ and $I$ magnitudes and $V-I$ color of the source assuming a bar angle of 25 degrees \citep{nataf13} and that the source star is located in the bar.
First cite the first one \citep{kervella04b} and then cite the second one \citep{kervella04a}.

\begin{deluxetable*}{lcccc}
\tablecaption{Limb-darkening Coefficients}
\tablewidth{0pt}
\tablehead{
\multicolumn{1}{c}{Event}              & 
\multicolumn{1}{c}{$\Gamma_{V}$}                &
\multicolumn{1}{c}{$\Gamma_{R}$}                &
\multicolumn{1}{c}{$\Gamma_{I}$}                &
\multicolumn{1}{c}{$\Gamma_{N}$}
} 
\startdata
OGLE-2004-BLG-368    &   0.73   &   0.63   &   0.52   &   0.58   \\
MOA-2005-BLG-36      &   0.80   &   0.71   &   0.59   &   0.65   \\
OGLE-2012-BLG-0211   &   0.57   &   0.49   &   0.41   &   0.45   \\
OGLE-2012-BLG-0456   &    ...   &   0.64   &   0.53   &   0.59   \\
MOA-2012-BLG-532     &   0.64   &   0.55   &   0.46   &   0.50   \\
MOA-2013-BLG-029     &    ...   &   0.61   &   0.51   &   0.56
\enddata
\tablecomments{
Here "$N$" indicates that the data are unfiltered.
}
\label{tab:parms_ld}
\end{deluxetable*}

Because finite-source effects are present in all events, variations in the surface brightness of the source due to limb-darkening must be accounted for to correctly compute the magnification.
To do this, we model the surface brightness profile as $S_{\lambda} \propto 1-\Gamma_{\lambda}(1-3{\rm cos}(\phi)/2)$, where $\Gamma_{\lambda}$ is the bandpass-specific limb-darkening coefficient and $\phi$ is the angle between the line of sight toward the source and the normal to the source surface.
We use the de-reddened color and brightness of the source star to estimate its spectral type, surface temperature $T_{\rm eff}$, and surface gravity log $g$ from the calibration tables of \citet{cox00}.
Using $T_{\rm eff}$ and log $g$ we interpolate across the tables of \citet{claret00} to obtain $\Gamma_{\lambda}$, assuming solar metallicity and a microturbulent velocity of $v = 2$ km s$^{-1}$.
Table \ref{tab:parms_ld} lists the bandpass-specific limb-darkening coefficients used for each event.

\begin{deluxetable*}{lcccccccc}
\tablecaption{Best-fit Lensing Parameters}
\tablewidth{0pt}
\tablehead{
\multicolumn{1}{c}{Event}            &
\multicolumn{1}{c}{$\chi^{2}$/dof}   &
\multicolumn{1}{c}{$t_{0}$}          &
\multicolumn{1}{c}{$u_{0}$}          &
\multicolumn{1}{c}{$t_{\rm E}$}      &
\multicolumn{1}{c}{$\rho$}           &
\multicolumn{1}{c}{$s$}              &
\multicolumn{1}{c}{$q$}              &
\multicolumn{1}{c}{$\alpha$}         \\
\multicolumn{1}{c}{}                 &
\multicolumn{1}{c}{}                 &
\multicolumn{1}{c}{[days]}           &
\multicolumn{1}{c}{}                 &
\multicolumn{1}{c}{[days]}           &
\multicolumn{1}{c}{}                 &
\multicolumn{1}{c}{}                 &
\multicolumn{1}{c}{}                 &
\multicolumn{1}{c}{}
} 
\startdata
OGLE-2004-BLG-368    &   3488.2/3446   &   3185.240      &    0.041          &   5.3          &   0.040          &   2.27          &   0.66         &   4.912         \\
                     &                 &   $\pm$0.004    &    $\pm$0.001     &   $\pm$0.1     &   $\pm$0.002     &   $\pm$0.04     &   $\pm$0.03    &   $\pm$0.006    \\
MOA-2005-BLG-36      &   692.4/684     &   3550.680      &    0.0146         &   12.6         &   0.033          &   ...           &   ...          &   ...           \\
                     &                 &   $\pm$0.003    &    $\pm$0.0009    &   $\pm$0.5     &   $\pm$0.002     &   ...           &   ...          &   ...           \\
OGLE-2012-BLG-0211   &   2613.6/2603   &   6013.1185     &   -0.00015        &   49           &   0.00053        &   ...           &   ...          &   ...           \\
                     &                 &   $\pm$0.0003   &    $\pm$0.00002   &   $\pm$2       &   $\pm$0.00002   &   ...           &   ...          &   ...           \\
OGLE-2012-BLG-0456   &   5636.7/5593   &   6047.101      &    0.2769         &   7.356        &   0.04624        &   1.0684        &   0.944        &   2.2766        \\
                     &                 &   $\pm$0.006    &    $\pm$0.0003    &   $\pm$0.004   &   $\pm$0.00002   &   $\pm$0.0006   &   $\pm$0.004   &   $\pm$0.0008   \\
MOA-2012-BLG-532     &   6378.8/6325   &   6151.734      &    0.0135         &   13.3         &   0.00144        &   6.1           &   0.47         &   2.106         \\
                     &                 &   $\pm$0.002    &    $\pm$0.0002    &   $\pm$0.1     &   $\pm$0.00001   &   $\pm$0.1      &   $\pm$0.04    &   $\pm$0.008    \\
MOA-2013-BLG-029     &   7815.9/7744   &   6353.00       &    0.413          &   11.74        &   0.0209         &   0.894         &   0.051        &   3.781         \\
                     &                 &   $\pm$0.02     &    $\pm$0.007     &   $\pm$0.08    &   $\pm$0.0004    &   $\pm$0.002    &   $\pm$0.001   &   $\pm$0.003
\enddata
\tablecomments{
Here $t_{0}$ is listed as HJD-2450000.
}
\label{tab:parms_bestfit}
\end{deluxetable*}

\begin{figure}[h]
\centerline{
\includegraphics[width=9cm]{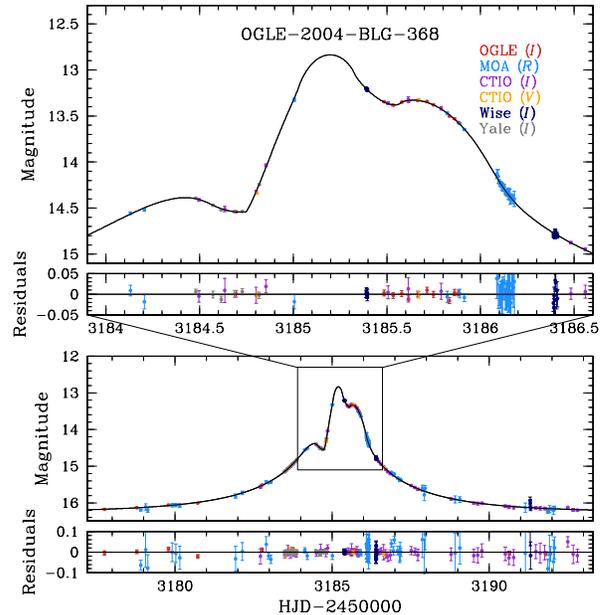}
}
\caption{
Light curve of OGLE-2004-BLG-368.
The telescope labels identify the color of the corresponding data and the bandpass.
The upper panel enhances the region where finite-source effects are prominent.
}
\label{fig:ob04368_lc}
\end{figure}

\begin{figure}[h]
\centerline{
\includegraphics[width=9cm]{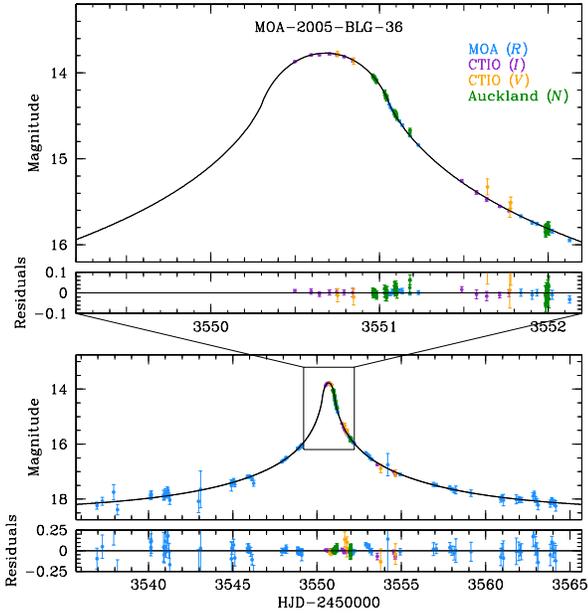}
}
\caption{
Light curve of MOA-2005-BLG-036.
The notations are the same as those in Figure \ref{fig:ob04368_lc}.
}
\label{fig:mb0536_lc}
\end{figure}

\begin{figure}[h]
\centerline{
\includegraphics[width=9cm]{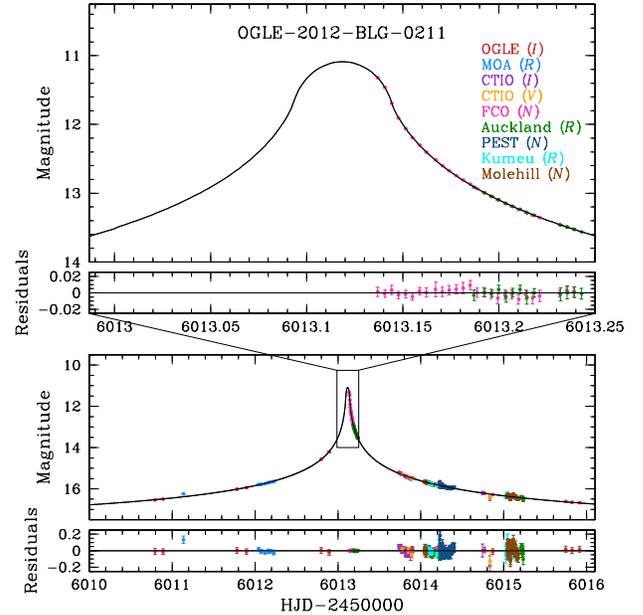}
}
\caption{
Light curve of OGLE-2012-BLG-0211.
The notations are the same as those in Figure \ref{fig:ob04368_lc}.
}
\label{fig:ob120211_lc}
\end{figure}

\begin{figure}[h]
\centerline{
\includegraphics[width=9cm]{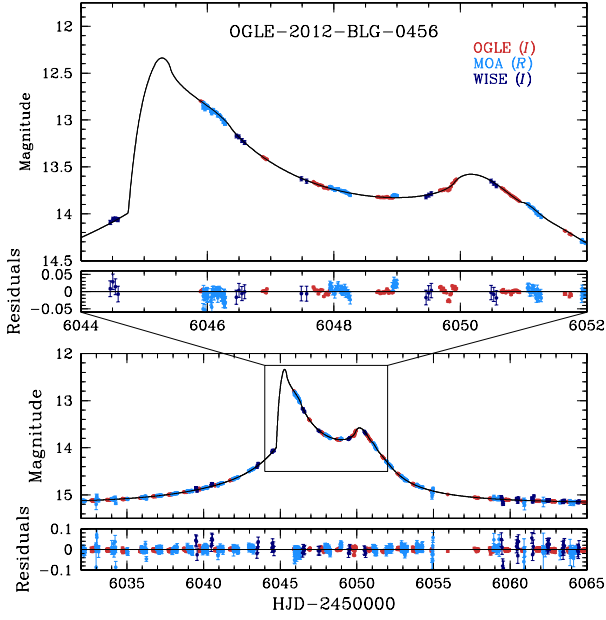}
}
\caption{
Light curve of OGLE-2012-BLG-0456.
The notations are the same as those in Figure \ref{fig:ob04368_lc}.
}
\label{fig:ob120456_lc}
\end{figure}

\begin{figure}[h]
\centerline{
\includegraphics[width=9cm]{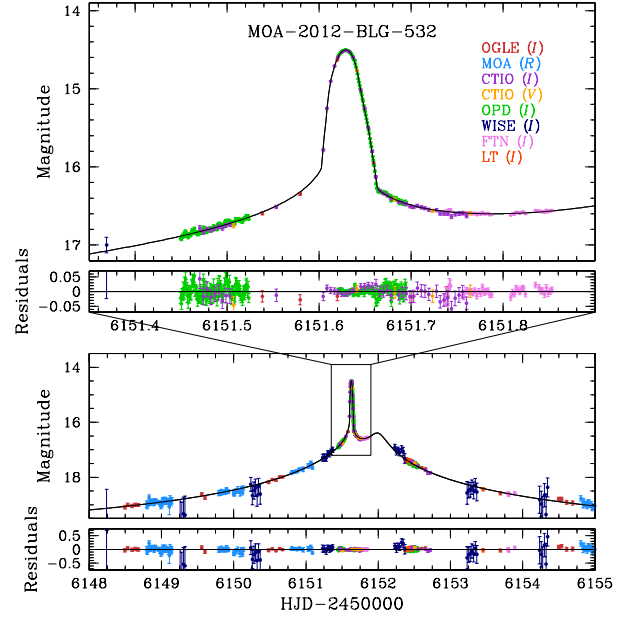}
}
\caption{
Light curve of MOA-2012-BLG-532.
The notations are the same as those in Figure \ref{fig:ob04368_lc}.
}
\label{fig:mb12532_lc}
\end{figure}

\begin{figure}[h]
\centerline{
\includegraphics[width=9cm]{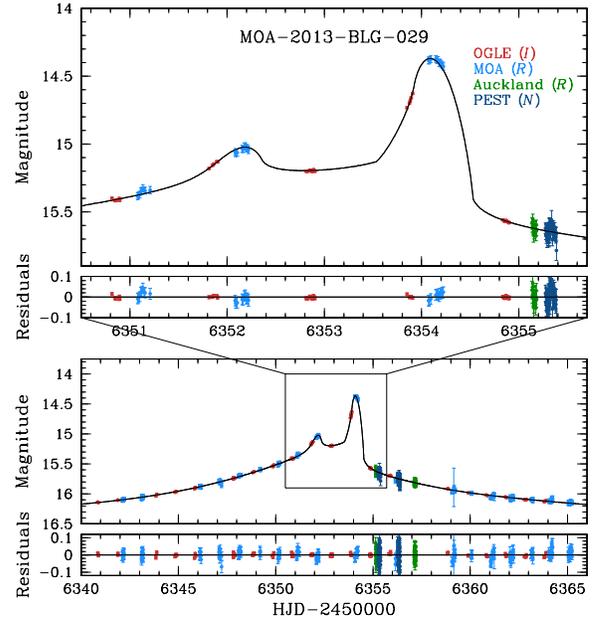}
}
\caption{
Light curve of MOA-2013-BLG-029.
The notations are the same as those in Figure \ref{fig:ob04368_lc}.
}
\label{fig:mb13029_lc}
\end{figure}

Table \ref{tab:parms_bestfit} gives the best-fit lensing parameters and their uncertainties for each event.
We note that for both OGLE-2012-BLG-0211 and MOA-2012-BLG-532, the source is quite faint.
In such cases, the flux measurement uncertainties can be of-order the flux measurements themselves and systematics in the baseline data can thus introduce biases in the determination of {\te}.
For a lengthier discussion of the effect such systematics can have on the microlensing parameters derived from the light curve, please see \citet{yee12}.
We also include the measurements of the source and blend flux, $F_{s}$ and $F_{b}$, respectively, the ratio $F_{b}/F_{s}$, and their uncertainties in Table \ref{tab:parms_fbfs}.
We restrict ourselves to those events with OGLE data, as their reported magnitudes are approximately calibrated, and thus quote these values only for the OGLE data.
These parameters are useful when measuring the relative lens-source proper motion and lens flux via the methods described in \citet{bennett07}.
In Figures \ref{fig:ob04368_lc} -- \ref{fig:mb13029_lc} we present the light curves of the individual events, superimposing the light curve of the best-fit model.

\begin{deluxetable*}{lccc}
\tablecaption{Source and Blend Fluxes}
\tablewidth{0pt}
\tablehead{
\multicolumn{1}{c}{Event}                & 
\multicolumn{1}{c}{$F_{s}$}              &
\multicolumn{1}{c}{$F_{b}$}              &
\multicolumn{1}{c}{$F_{b}/F_{s}$}
}
\startdata
OGLE-2004-BLG-368    &   5.1 $\pm$ 0.1         &   -0.2 $\pm$ 0.1        &   -0.03 $\pm$ 0.02       \\
OGLE-2012-BLG-0211   &   0.149 $\pm$ 0.007     &   0.740 $\pm$ 0.007     &   5.0 $\pm$ 0.3              \\
OGLE-2012-BLG-0456   &   13.53 $\pm$ 0.01      &   -0.07 $\pm$ 0.01      &   -0.0053 $\pm$ 0.0007   \\
MOA-2012-BLG-532     &   0.1016 $\pm$ 0.0009   &   0.0293 $\pm$ 0.0006   &   0.29 $\pm$ 0.01        \\
MOA-2013-BLG-029     &   4.21 $\pm$ 0.08       &   0.24 $\pm$ 0.08       &   0.06 $\pm$ 0.02
\enddata
\tablecomments{
The values quoted here are for a photometric system in which one photon per second is measured from an 18th-magnitude object, and are measured from the OGLE data.
}
\label{tab:parms_fbfs}
\end{deluxetable*}

The event OGLE-2004-BLG-368 was produced by a binary lens with a projected separation greater than the Einstein radius.
In the case of a wide binary with $s>1$ there exist two caustics, one close to each lens component.
The perturbation occurred when the source passed over the caustic nearest the more massive lens component.
The first bump at $\rm {HJD'} = {\rm HJD} - 2450000 \approx 3184.4$ was produced by the source's approach close to a cusp of the caustic.
After the initial bump, the center of the source crossed the caustic 4 times, but $\rho$ is larger than the gaps between the caustic entrance and exit and thus each of the two pairings of a caustic entrance and exit appears as a single bump, the first at ${\rm HJD'} \approx 3185.2$ and the second at ${\rm HJD' \approx 3185.7}$.

MOA-2005-BLG-36 was caused by a single lensing mass.
This event occurred in a highly extinguished region, so while the source has a faint apparent magnitude, it is actually a giant.
The impact parameter is smaller than the normalized source radius, $u_{\rm 0} < \rho$, indicating that the lens passed over the source star.
As a result, the peak of the light curve is affected by severe finite-source effects and deviates from the standard point-source model.

OGLE-2012-0211 was also caused by a single lensing mass.
As in the case of MOA-2005-BLG-36, $u_{\rm 0} < \rho$, so the peak of the light curve is similarly affected by finite-source effects.
Although the coverage of the peak is incomplete, the data sufficiently cover the portion of the trajectory during which the lens finishes traversing the disk of the source, allowing for the precise measurement of $\rho$.

The event OGLE-2012-BLG-0456 was produced by a binary with a projected separation near unity.
Such a resonant lens configuration creates a single large caustic that is located between the two lens masses.
The initial bump in the light curve occurred when the source crossed the cusp near the less-massive lens component at ${\rm HJD'} \approx 6045$.
The source then passed inside the caustic, producing the second bump when it exited the caustic at ${\rm HJD'} \approx 6050$.
Due to the severe finite-source effects that arise from $\rho$ being comparable to the size of the caustic features, each caustic crossing did not form the sharp spike that is characteristic of a point-like source, but instead exhibited a rounded shape.

The light curve of MOA-2012-BLG-532 is characterized by a strong bump followed by a weaker bump near the peak of the light curve.
From modeling we find that this central deviation can be explained either by a very close, $s \ll 1$, or a very wide, $s \gg 1$, binary.
This degeneracy, known as the close/wide binary degeneracy, is caused by the similarity in both size and shape of the central caustic of the two sets of binary configurations \citep{griest98,dominik99}.
We find that the wide solution yields a slightly better fit by $\Delta\chi^{2} \approx 10$ and thus present the analysis based on the wide solution.
In this case, the close/wide degeneracy stems from the intrinsic similarity of the caustics and thus does not lead to a significant difference in the characteristic length or time scales.
In the case of MOA-2012-BLG-532, the values for {\thetae} and $\mu$ for the close and wide solutions are within one sigma of each other.
We find that the strong bump was caused by the passage of the source across the tip of the protruding cusp and that the weak bump was produced by the source's approach to another cusp of the central caustic.

Similar to MOA-2012-BLG-532, the light curve of MOA-2013-BLG-029 is characterized by two bumps near the peak.
We find that the event was produced by a close binary with a separation not much smaller than the Einstein radius.
As with OGLE-2012-BLG-0456, this produces a single, resonant caustic.
In this case, the resonant caustic is elongated, and the deviations were due to successive passages of the source near the cusps of the elongated portion of the caustic.
We note that the mass ratio of the binary lens, $q \approx 0.05$, is small.
Considering that the time scale of the event, $t_{\rm E} \approx 12$ days, is typical for events produced by low-mass stars, the small $q$ suggests that the lower-mass lens component is likely a substellar object such as a brown dwarf or even a giant planet.

\begin{figure*}
\centerline{
\includegraphics[width=12cm]{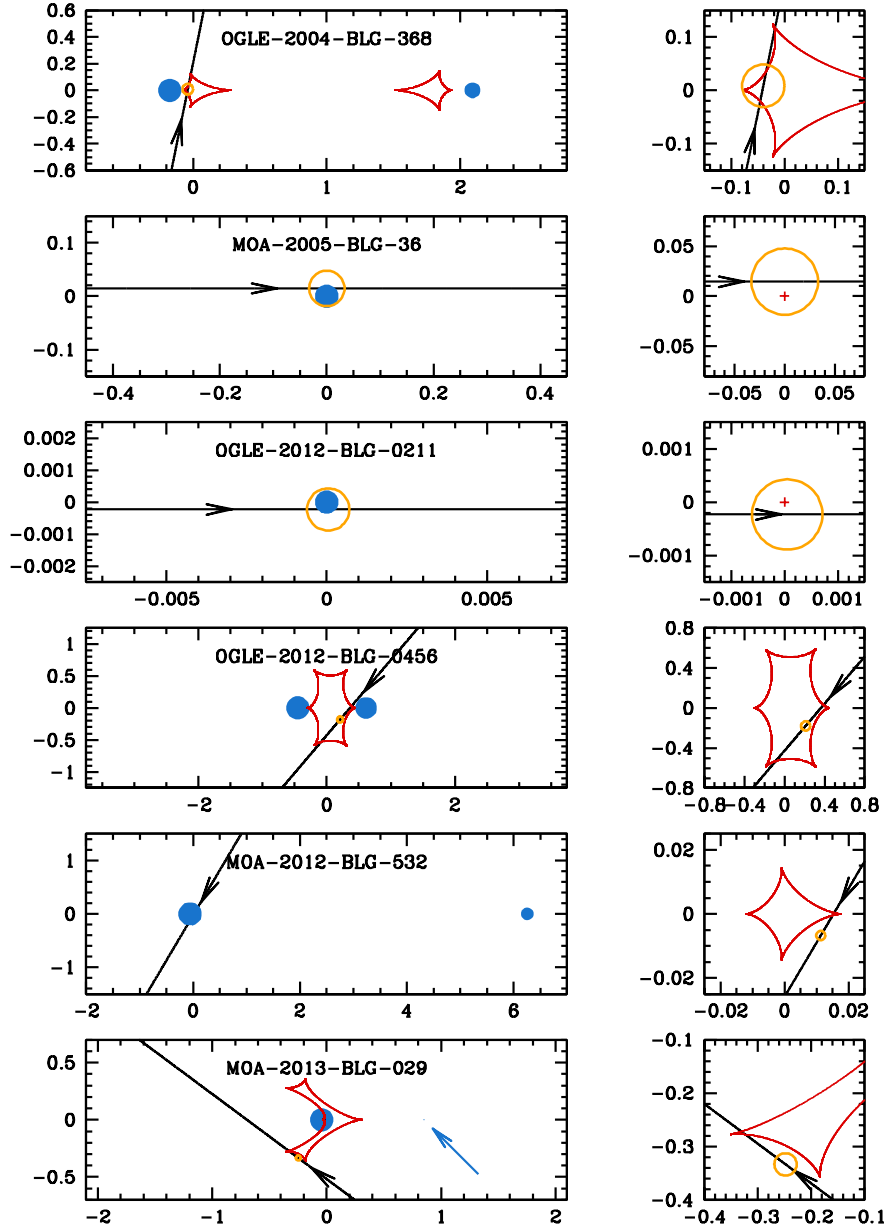}
}
\caption{
The geometry of each lensing event.
The blue filled circles mark the positions of the lens masses, with arbitrary absolute sizes but relative sizes scaled according to $q$.
The orange empty circle denotes the source.
The red lines show the caustic curves.
For MOA-2005-BLG-36 and OGLE-2012-BLG-0211, the two single-lens events, the red plus in the right panel marks the position of the lens at the origin.
The black line and arrow mark the source trajectory.
All lengths are normalized by the angular Einstein radius corresponding to the total mass of the lens system.
}
\label{fig:geometry}
\end{figure*}

Figure \ref{fig:geometry} contains the geometry of each event, showing the source trajectory with respect to the lens position and caustics
(in the case of the binary lenses).
For all events we find no strong evidence for the presence of higher-order effects such as microlens parallax caused by the change of the observer's position induced by the Earth's orbital motion \citep{gould92a,alcock95} or a change in the lens position induced by the orbital motion of the lens \citep{dominik98,albrow00,shin11,skowron11,jung13}.
This is unsurprising given the relatively short time scale of most of the events, with ${\te} \lesssim 13$ days.
The time scale of OGLE-2012-BLG-0211 is longer than those of the other events, but it is not long enough for secure detection of the parallax effect.
Additionally, the event is caused by a single mass and thus there is no orbital motion of the lens.
We did not consider the effects of terrestrial parallax \citep{hardy95,holz96,gould09}.

\section{Results} \label{sec:results}

In Table \ref{tab:parms_ppm} we present {\thetae} and $\mu$ for the events newly analyzed in this paper as well as the other events with high $\mu$ from the literature, listed together chronologically.
This catalog includes 20 lensing events over the period 2004 -- 2013 with high relative lens-source proper motion, $\mu \gtrsim 8$ mas yr$^{-1}$, providing a list of events whose source and lens will be sufficiently separated in the next $\lesssim$12 years for follow-up observations using ground-based AO systems or space telescopes that will be able to resolve each component separately.

There exist several instruments that are currently able to observe the Galactic Bulge and achieve the spatial resolution necessary to resolve the lens from the source for each of these events ($\lesssim$0.1$\arcsec$).
The {\it Hubble Space Telescope} ({\it HST}) obtains $\sim$0.1$\arcsec$ resolution in the NIR with WFC3 \citep{sabbi13}.
Additionally, imaging in the optical with {\it HST} provides even greater spatial resolution ($\sim$0.04$\arcsec$) and allows for direct comparison to OGLE $I$-band photometry.
On the ground there are multiple instruments with AO systems able to obtain angular resolution at or near the diffraction limit of their respective telescopes in the NIR, including NaCo on the 8.2m UT4 VLT at Cerro Paranal in Chile \citep{lagrange03}, ALTAIR on the 8.1m Gemini N \citep{christou10} and IRCS on the 8.2m Subaru Telescope \citep{kobayashi00}, both at Mauna Kea in Hawaii, GeMs on the 8.1m Gemini S at Cerro Pachon in Chile \citep{rigaut12}, and NIRC2 on the 10m Keck II telescope at Mauna Kea in Hawaii \citep{wizinowich00}.

There is also a myriad of planned next-generation space and ground-based telescopes that will be able to achieve such high angular resolution.
In space there will be Euclid, which will achieve $\sim$0.1$\arcsec$ and $\sim$0.3$\arcsec$ resolution at visible and NIR wavelengths, respectively \citep{laureijs11}, JWST, which will achieve $\sim$0.03$\arcsec$ resolution in the NIR with NIRCam \citep{greene10}, and, potentially, WFIRST, whose AFTA design can achieve $\sim$0.1$\arcsec$ resolution in the NIR \citep{spergel13}.
Currently there are three extremely large telescopes with diameters greater than 20 meters planned, each with a proposed AO system to be used with a NIR imager, including EPICS on the 39.3m European Extremely Large Telescope \citep{verinaud10}, NFIRAOS on the 30m Thirty Meter Telescope \citep{ellerbroek13}, and GMTIFS on the 25.4m Giant Magellan Telescope \citep{mcgregor12}.

\begin{deluxetable*}{lccl}
\tablecaption{Einstein Radius and Proper Motion}
\tablewidth{0pt}
\tablehead{
\multicolumn{1}{c}{Event}              & 
\multicolumn{1}{c}{$\theta_{\rm E}$}   &
\multicolumn{1}{c}{$\mu$}              &
\multicolumn{1}{c}{Reference}          \\
\multicolumn{1}{c}{}                   &
\multicolumn{1}{c}{[mas]}              &
\multicolumn{1}{c}{[mas yr$^{-1}$]}    &
\multicolumn{1}{c}{}
} 
\startdata
OGLE-2004-BLG-035    &   1.4 $\pm$ 0.2           &   7.9 $\pm$ 0.7$^{a}$     &   \citet{shin12c}      \\
OGLE-2004-BLG-368    &   0.14 $\pm$ 0.01         &   9.7 $\pm$ 0.9           &   This work            \\
OGLE-2004-BLG-482    &   $\approx$0.4$^{b}$      &   $\approx$16$^{b}$       &   \citet{zub11}        \\
OGLE-2005-BLG-169    &   1.0 $\pm$ 0.2$^{c}$     &   8.4 $\pm$ 1.7$^{c}$     &   \citet{gould06}      \\
MOA-2005-BLG-36      &   0.30 $\pm$ 0.03         &   8.8 $\pm$ 0.8           &   This work            \\
OGLE-2006-BLG-277    &   1.4 $\pm$ 0.1           &   13 $\pm$ 1              &   \citet{park13}       \\
MOA-2007-BLG-146     &   0.44 $\pm$ 0.04$^{d}$   &   10.2 $\pm$ 0.9$^{d}$    &   \citet{shin12a}      \\
OGLE-2007-BLG-224    &   0.91 $\pm$ 0.04         &   48 $\pm$ 2              &   \citet{gould09}      \\
MOA-2007-BLG-400     &   0.32 $\pm$ 0.02         &   8.2 $\pm$ 0.5           &   \citet{dong09a}      \\
MOA-2010-BLG-477     &   1.4 $\pm$ 0.11          &   10.3 $\pm$ 0.8          &   \citet{bachelet12}   \\
MOA-2010-BLG-546     &   0.21 $\pm$ 0.03$^{e}$   &   8.4 $\pm$ 1.2$^{e}$     &   \citet{shin12a}      \\
MOA-2011-BLG-040     &   1.17 $\pm$ 0.01$^{d}$   &   8.04 $\pm$ 0.02$^{d}$   &   \citet{shin13}       \\
OGLE-2011-BLG-0417   &   2.44 $\pm$ 0.02         &   9.66 $\pm$ 0.07         &   \citet{shin12b}      \\
MOA-2011-BLG-262     &   $\approx$0.23$^{b,f}$   &   $\approx$19.6$^{b,f}$   &   \citet{skowron13}    \\
MOA-2011-BLG-274     &   0.08 $\pm$ 0.01         &   11.2 $\pm$ 1.0          &   \citet{choi12}       \\
OGLE-2012-BLG-0211   &   1.6 $\pm$ 0.2           &   12 $\pm$ 1              &   This work            \\
OGLE-2012-BLG-0456   &   0.23 $\pm$ 0.02         &   12 $\pm$ 1              &   This work            \\
MOA-2012-BLG-532     &   0.34 $\pm$ 0.02$^{e}$   &   9.1 $\pm$ 0.8$^{e}$     &   This work            \\
MOA-2013-BLG-029     &   0.27 $\pm$ 0.02         &   8.4 $\pm$ 0.8           &   This work            \\
MOA-2013-BLG-220     &   0.45 $\pm$ 0.03         &   12.5 $\pm$ 1            &   \citet{yee14}
\enddata
\tablenotetext{a}{We compute $\mu$ using the values of {\thetae} and {\te} quoted in the paper for the fit with the lowest $\chi^{2}$, adding the uncertainties in quadrature.}
\tablenotetext{b}{The values for {\thetae} and $\mu$ are taken from the text, but no uncertainties are provided.}
\tablenotetext{c}{The quoted uncertainties are 3$\sigma$.}
\tablenotetext{d}{The values quoted here are for the close solution, which is the model with the lowest $\chi^{2}$.}
\tablenotetext{e}{The values quoted here are for the wide solution, which is the model with the lowest $\chi^{2}$.}
\tablenotetext{f}{The paper includes the values $\theta_{\rm E} = 0.14$ mas and $\mu = 11.6$ mas yr$^{-1}$ for a competing but nearly equally likely model, which would still cause it to be included in our sample.}
\label{tab:parms_ppm}
\end{deluxetable*}

\section{Conclusions} \label{sec:conclusions}

Here we have presented a catalog of microlensing events over the period 2004 -- 2013 with high proper motion, $\mu \gtrsim 8$ mas yr$^{-1}$.
In the next $\lesssim$12 years, each of the events in our catalog will be sufficiently separated for direct imaging of the lens system.
There are several ground-based telescopes with AO systems operating at or near the diffraction limit and space telescopes that have or will have the angular resolution necessary to resolve the lens from the source on that time scale.
Such observations will further characterize each lens system and provide valuable insight into Galactic structure.

We also urge for immediate high-resolution images to be taken of the six events whose lens and source are  currently $\lesssim$20 mas.
This includes MOA-2011-BLG-040 ($\sim$24 mas), OGLE-2012-BLG-0211 ($\sim$24), OGLE-2012-BLG-0456 ($\sim$24), MOA-2012-BLG-532 ($\sim$18), MOA-2013-BLG-029 ($\sim$8.4), and MOA-2013-BLG-220 ($\sim$12.5).
For these events, the lens and source will still be unresolved in high-resolution images taken in the very near future.
It will then be possible to compare the estimate of the lens flux obtained using this initial image with that obtained directly using a future high-resolution image taken when the source and lens are resolved.
This comparison will provide an empirical check on the robustness of the single-epoch method currently being used to estimate lens masses for many events.

\acknowledgments
Work by C.~B.~Henderson is generously supported by a National Science Foundation Graduate Research Fellowship under grant No.~DGE-0822215.
T.~S.~is supported by the grant JSPS23340044 and JSPS24253004.
Work by C.~H.~was supported by the Creative Research Initiative Program (2009-0081561) of the National Research Foundation of Korea.
A.~G.~and B.~S.~G.~acknowledge support from NSF AST-1103471 and from NASA grant NNX12AB99G.
S.~D.~was supported through a Ralph E.~and Doris M.~Hansmann Membership at the IAS and NSF grant AST-0807444.
Work by J.~C.~Y.~was performed in part under contract with the California Institute of Technology (Caltech) funded by NASA through the Sagan Fellowship Program.
DMB, MD, KH, MH, CS, RAS, and YT acknowledge grant NPRP-09-476-1-78 from the Qatar National Research Fund (a member of the Qatar Foundation).

\bibliographystyle{apj}

\end{document}